\documentclass[9pt,twocolumn,twoside]{opticajnl}
\journal{opticajournal} 

\setboolean{shortarticle}{true}

\usepackage{xcolor}

\title{Closed-Loop Polarization Mode Dispersion Mitigation for Fibre-Optic Time and Frequency Transfer}

\author[1,*]{Thomas Fordell}
\author[1]{Anders E. Wallin}
\author[1]{Thomas Lindvall}
\author[1]{Kalle Hanhijärvi}

\affil[1]{VTT Technical Research Centre of Finland Ltd, National Metrology Institute VTT MIKES, P.O.\ Box 1000, FI-02044 VTT, Finland}

\affil[*]{thomas.fordell@vtt.fi}

\begin{abstract}
A polarization switching pulse interleaver is shown to be effective in reducing timing noise due to polarization mode dispersion in time and frequency transfer based on mode-locked lasers {\color{black} and standard single-mode fibres}. In closed-loop time transfer over a 30-km dispersion-compensated fibre link with 300~fs of differential group delay, polarization interleaving reduced the delay variations to <~20~fs. The results indicate that the remaining drift is  caused by polarization-dependent loss and by AM-to-PM noise conversion in the photodiodes, suggesting the need for a  "double-balanced" phase detector in the receiver, that is, a phase detector balanced in power and polarization. {\color{black} By mitigating the polarization dependence, this work demonstrates a simple approach that can potentially yield sub-femtosecond-level, long-term time transfer in long-haul fibre links utilizing standard single-mode fibres. 

}

\end{abstract}

\setboolean{displaycopyright}{false} 

\begin{document}

\maketitle

Accurate synchronization of remote systems and event timing is an enabler of many large-scale scientific experiments. Global-navigation satellite systems can, in post processing, provide sub-nanosecond level accuracy against Coordinated Universal Time (UTC) globally while fibre-optic links can provide several orders of magnitude lower synchronization error albeit over more limited distances. Free-space optical links based on frequency combs can yield even attosecond level line-of-sight timing {\color{black}{over several 100~km}} \cite{Caldwell2023a}, but polarization mode dispersion (PMD)  is likely to become a considerable obstacle for sub-femtosecond long-haul fiber-optic links {\color{black} that utilize  standard single-mode (SM) fibres} \cite{Xu2021a, Zhang2017a, Gibbon2015a}. Impressive results have been obtained with long-haul optical frequency transfer in fibres (e.g. \cite{Droste2013a, Bercy2014a,  Guillou-Camargo2018a, Schioppo2022a}). It should be noted, however, that time transfer (synchronization) is quite different from optical frequency transfer. For instance, whereas a step change in phase will average down in frequency transfer, it will result in a permanent error in time transfer. A frequency offset, however small, will result in an ever increasing synchronization error, and this will  
 not influence Allan deviation nor Time deviation, the two often presented statistical measures of link performance. Furthermore, unlike optical frequency transfer, time transfer may require a considerable amount of optical dispersion compensation, although dispersion compensation can, in principle, be carried out digitally in a coherent receiver \cite{Xie2011}.  Spooled, dispersion-compensating fibre (DCF) can have a differential group delay of several 100~fs. 
   Chirped fibre-bragg gratings are an alternative to DCF but they too can have significant {\color{black}differential group delay} \cite{Ning2005a}.
\begin{figure}[t]
\centering
\includegraphics[width=\linewidth]{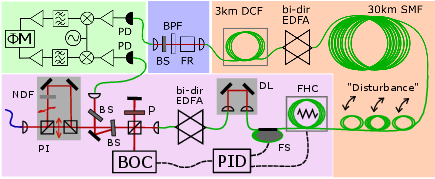}
\caption{Experimental setup. Light from the frequency-comb enters via the blue PM  fibre (lower left corner) with a polarization 45$^o$ compared to the axis of the first polarizing beam splitter cube. PI: polarization-switching pulse interleaver; NDF: neutral density filter; BS: beam splitter; P: polarizer; BOC: balanced optical cross-correlator (see Fig. 2); PID: proportional, integral, derivative controller; DL: free-space delay line; FS: fiber stretcher; FHC: fibre heater/cooler; DCF: dispersion-compensating fibre module; FR: faraday rotator; BPF: 12-nm bandpass filter at 1560~nm; $\Phi$M: phase meter.}
\label{fig:setup}
\end{figure}
 PMD can be avoided using polarization maintaining fibres \cite{Xin2014a, Xin2017a, Xin2018b}, but the relatively higher loss and higher cost is problematic for long distance links. Also, the ability to use pre-existing telecommunications networks is often desirable.

In any ultra-stable link, the instabilities of the transfer medium must be compensated for. This can be achieved passively (two-way transfer, e.g. \cite{He2013a, Bercy2014a,Deschenes2016a,Sinclair2019a, Abuduweili2020a, Schioppo2022a}) or actively by having the receiver reflect part of the light back so that the transmitter can compare the incoming phase to the outgoing and form a phase-locked loop (PLL). In the latter case, a Faraday rotator is often used so that upon return to the transmitter, the polarization is rotated 90~degrees compared to the outgoing light, enabling reliable phase detection and PLL operation. 
 The Faraday rotator makes the round-trip time polarization independent as long as the round-trip time is much smaller than the PMD noise time scale; therefore, even though the forward and backward signals suffer from PMD, the transmitter is unaware of it and cannot correct for it. A piezo-electric polarization scrambler can be used (e.g. \cite{Yao2002a}) but even a fast scrambler ($\sim$MHz) can yield noise close to the carrier. Microwave polarization modulation on a great circle of the Poincare sphere can be realized with a single electro-optic phase modulator \cite{Kersey1990a}. For example, in \cite{Zhang2014a} polarization modulation at 4.6~GHz was used in an RF-over-fibre link to make the signal at 2.4~GHz seem unpolarized to the receiver.

In \cite{Fordell2022a} a simple solution for mode-locked-laser-based systems in the form of a polarization-switching pulse interleaver was put forward and shown to significantly reduce PMD noise in an open-loop proof-of-concept configuration. {\color{black} Briefly, the temporal position $\langle t \rangle$ of an optical pulse is given by $\langle t \rangle = \langle E|t|E\rangle/\langle E|E\rangle$, where $|E\rangle = (E_x(t), E_y(t))$ is the Jones vector. If the medium is linear and without polarization dependent loss, then upon propagation, the Fourier transform of the Jones vector is transformed according to $|\tilde{E}_2\rangle = BU|\tilde{E}_1\rangle$, where the scalar $B$ contains polarization independent losses and dispersion and the unitary matrix $U$ contains polarization dependent effects.  It can then be shown that the average arrival time of two orthogonal pulses $|E_{11}\rangle$ and $|E_{12}\rangle$ is polarization independent \cite{Fordell2022a}. This means that in a polarization-interleaved pulse train, such as used in this work, only two consecutive pulses are needed, translating into a very high bandwidth for the PMD cancellation. Direct photodetection and phase measurement at a harmonic of the pulse frequency, as done here, is a straightforward and "automatic" way of performing the pulse averaging; however, as will be shown below, there are problems with this approach.}

The present work demonstrates that {\color{black}polarization switching} can be successfully operated in closed-loop and over a 30-km SM fiber link with an additional 3-km of fibre in a DCF module. The 30-km link distance was chosen due to available fiber spools considering that the link should be dispersion compensated to a very high degree. The DCF module has a {\color{black}differential group delay} of 100~fs and the {\color{black}entire link (30~km + 3~km)} about 300~fs. The various components are not spliced together but connectorized, and the free-space-to-fibre couplings are not mode-matched properly, which results in a total one-way loss equal to $\sim$100~km of SM fibre.

\begin{figure}[t]
\centering
\includegraphics[width=0.3\linewidth]{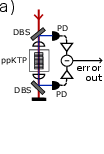} 
\includegraphics[width=0.69\linewidth]{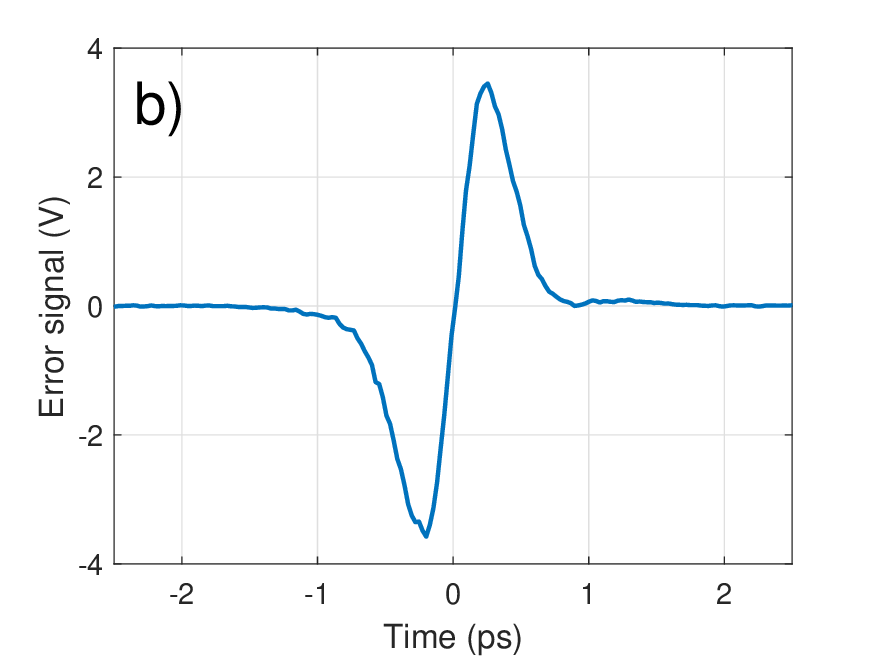}
\caption{a) In-loop phase detector, which is based on a temperature-controlled 3.15-mm-long ppKTP crystal \cite{Kim2007a, Kartner2007a}. PD: Si photodiode; DBS: dichroic beamsplitter.  b) Representative BOC error signal with an effective  transimpedance gain of 600~M$\Omega$ (detector bandwidth 1~kHz.). The signal amplitude depends on the settings of the EDFAs and on whether single (250~MHz) or dual polarization (500~MHz) pulse trains are used. Total power toward detector typically 0.5-1~mW.}
\label{fig:PM_and_BOC}
\end{figure}

The experimental setup is depicted in Fig.~\ref{fig:setup}. Light from an optically phase-locked frequency comb (MenloSystems FC1500-250-ULN) is filtered and amplified (bandwidth $\sim$6~nm at 1560~nm, pulse duration about 620~fs) and enters the setup via the blue polarization-maintaining (PM) fibre at the lower left corner. A polarization-switching pulse interleaver with a delay equal to half the pulse period increases the pulse repetition rate from 250~MHz to 500~MHz. The delay was adjusted based on the RF power spectrum and was not actively stabilized. A 10/90 beam splitter (BS) is used to sample the light into the reference arm of the out-of-loop phase meter, which operates by downmixing the 8th harmonic (2~GHz) of the laser to 10~MHz suitable for the Symmetricom 3120A phase noise probe ($\Phi$M).  For the in-loop phase detection, balanced detectors based on microwave mixers and optical second harmonic generation were tested. In terms of PMD mitigation, both produced similar results, but here results only from optical phase detection is presented since 
 the inline balanced optical cross-correlator (BOC, \cite{Kim2007a, Kartner2007a}) is {\color{black} particularly} attractive for long term operation due to its compact design, a schematic of which is shown in Fig.~\ref{fig:PM_and_BOC}a. Here a 3.15~mm-long PPKTP crystal (type II phase matching) was used. Higher sensitivities can be obtained with waveguides \cite{Callahan2014a, Safak2022a}.
 The BOC requires the pulses to enter the birefringent crystal with a small polarization-dependent delay; therefore, the use of alternating polarizations requires a compensating delay in the reference arm {\color{black}of the in-loop phase detector} (e.g. an identical PPKTP crystal); however, as explained earlier, due to the Faraday-rotator at the receiver, {\color{black}transmitting} a single polarization is enough {\color{black} (since round-trip time is polarization independent)} 
  and so a polarizer is simply put in the reference arm to select every other pulse for phase detection, which means that half the signal is lost. A representative error signal is shown in Fig.~\ref{fig:PM_and_BOC}b. 

Before entering the transmission line, the pulses pass through a bi-directional EDFA, a delay stage (DL), a piezo-electric fiber stretcher (FS) and a fiber heater/cooler (FHC), which {\color{black} consists of an aluminum frame with peltier elements along the circumference. The first (and only) fibre layer ($\sim$71~m) is wrapped along the surfaces of the peltiers', yielding a relatively short response time and enables heating as well as cooling. The present implementation provides a (driver-limited) phase tuning speed of $\sim$3~ps$/$s and a range of $\sim$50~ps (bandwidth $\sim$10~mHz). With more fibre layers and an improved driver these numbers could be increased significantly, but this was sufficient for the present work. The delay stage was only used for coarse adjustments, but it could also be controlled for achieving several nanoseconds of delay}. 
 A manual, 3-paddle polarization controller is used to add polarization noise, i.e., to simulate polarization variations that may occur in a real link on more or less any time scale. At the receiving end the pulses are amplified by another bi-directional EDFA and re-compressed by the DCF module. At the receiver the light is bandpass filtered (12~nm) and part of the light is directed to a high-power 20~GHz photodiode {\color{black}that provides the signal to the out-of-loop phase detector. The bias voltage of the photodiode} was carefully adjusted for each measurement in order to minimize AM-to-PM noise conversion \cite{Taylor2011a, Zhang2012a, Sun2014a}. Optimum bias voltages were found by modulating the pump power of EDFAs and minimizing the resulting phase drift. The reminder of the light is reflected back with orthogonal polarization due to the Faraday rotator. Back at the BOC, the pulse duration is $\sim 700$~fs. 
  Standard optomechanics and free-space optics were assembled on an optical table without enclosures or temperature stabilization. {\color{black} (Only the ppKTP crystal was stabilized at the optimum phase-matching temperature). As such, the setup} is not suitable for long-term studies on the femtosecond time scale.
\begin{figure}[t]
\centering
\includegraphics[width=\linewidth]{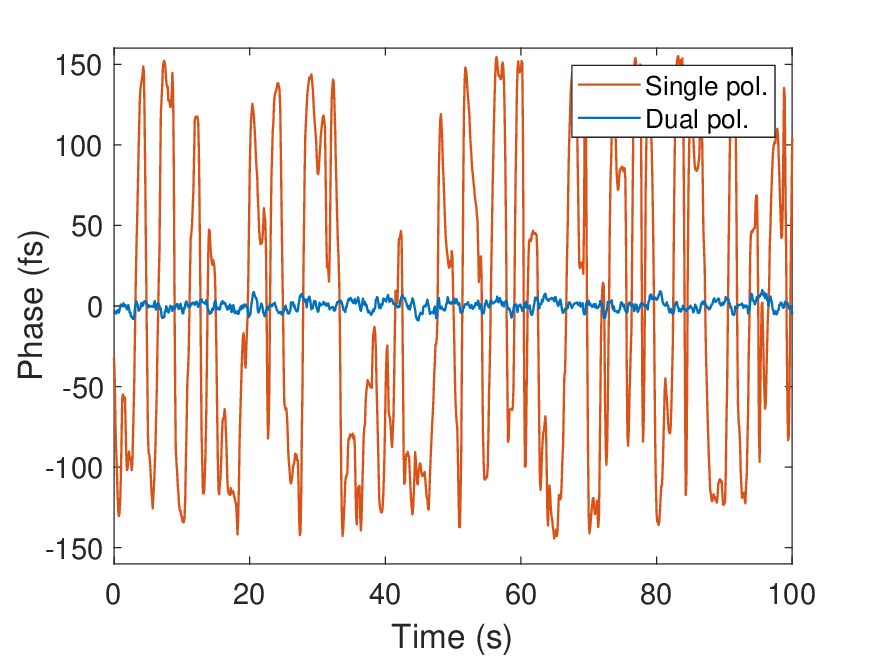}
\caption{Closed-loop microwave transfer (2~GHz) with single (red) and alternating orthogonal polarization (blue). During both measurements, the paddles on the 3-paddle polarization controller are continuously moved in a random pattern. Detection bandwidth was 5~Hz.} 
\label{fig:results1}
\end{figure}

With only a single polarization (delay arm blocked), the peak-to-peak phase variation is approximately 300~fs as the paddles of the polarization controller are manually adjusted continuously in a random pattern (Fig.~\ref{fig:results1}, red trace). With dual polarization, the peak-to-peak variation is reduced to <~20~fs. 

{\color{black}
Unsurprisingly, power fluctuations together with residual AM-to-PM conversion limits the short-term performance when the paddles are kept stationary. As seen from the Allan deviation plot in Fig. 4, where the phase noise level when the paddles are kept stationary, i.e., the system is free running (red trace), matches the estimated phase noise level due to power fluctuations (black dashed trace). The latter was obtained by determining the residual AM-to-PM coefficient $\kappa=\frac{\delta\Phi}{\delta P}$ by modulating the remote EDFA and recording the resulting phase deviation $\delta \Phi$. Power fluctuations $\delta P$ were determined by recording the varying bias current of the receiver photodiode using a bias-T placed after the photodiode and connected to an analog-to-digital converter. The estimate is then the Allan deviation of $\Phi(t) = \kappa P(t)$, where $P(t)$ is the power recorded during closed-loop operation.}
\begin{figure}[t]
\centering
\includegraphics[width=0.9\linewidth]{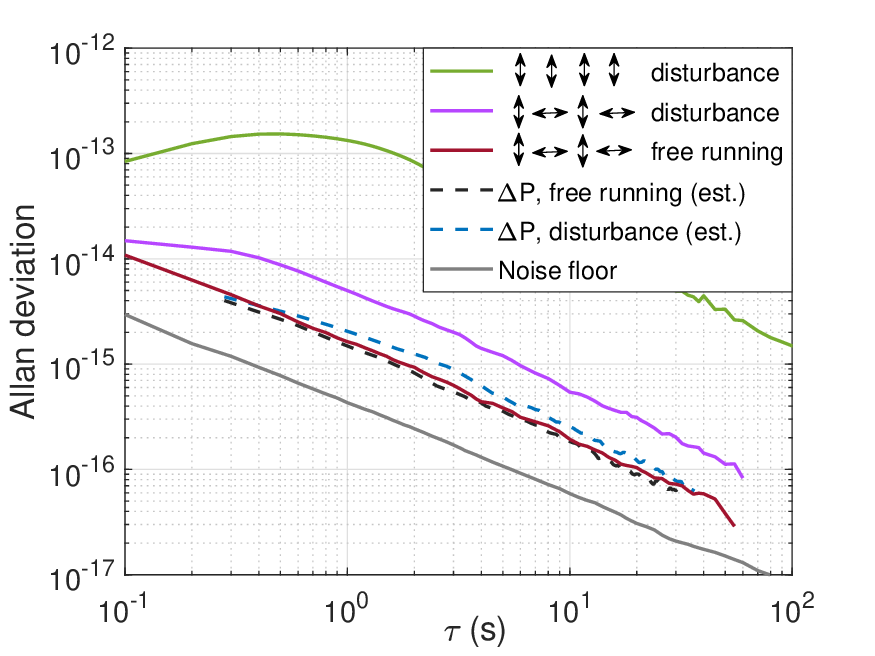}
\caption{Allan deviation for single polarization (green) and dual polarization (purple) with continuous, random movements of the paddles on the polarization controller. Dual polarization with paddles stationary is given in red. Dashed lines indicate estimated noise level based only on measured power fluctuations with (black) and without (blue) polarization disturbances. Detection bandwidth was 5~Hz.}
\label{fig:adev}
\end{figure}
\begin{figure}[t!]
\centering
\includegraphics[width=\linewidth]{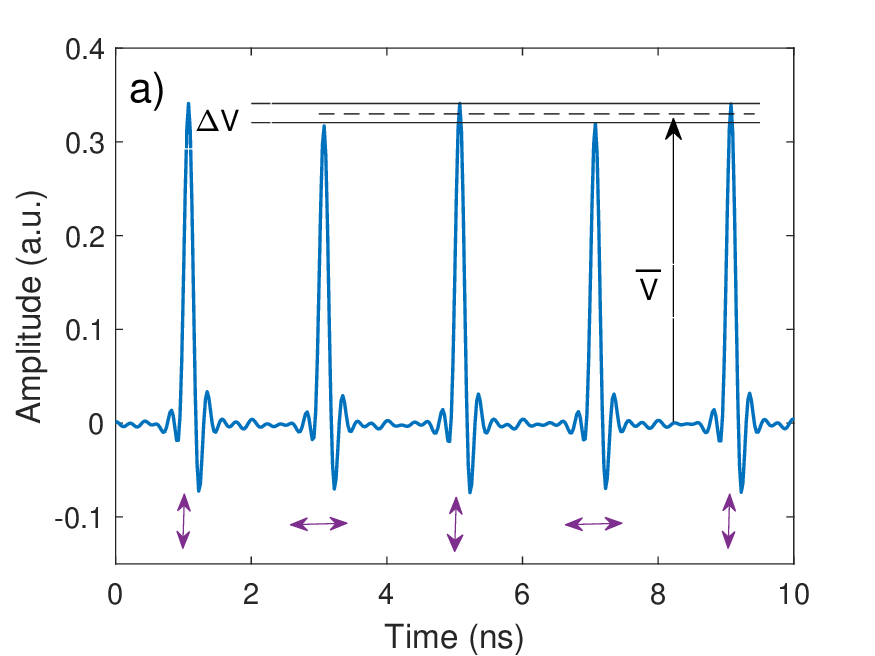}\\
\includegraphics[width=\linewidth]{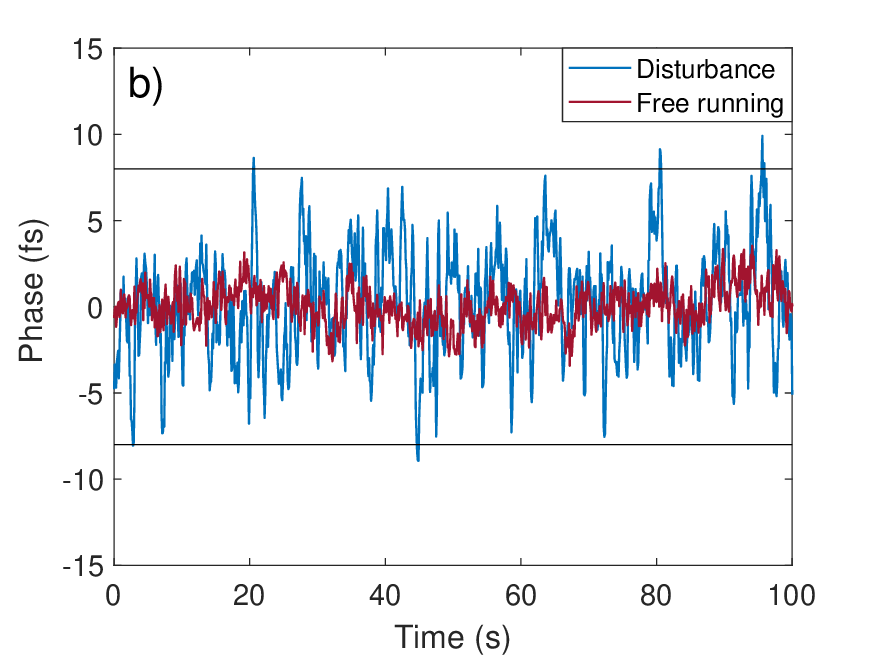}
\caption{a) Oscilloscope trace of the receiver photodiode signal (4~GHz bandwidth) when the polarization controller paddles are adjusted for maximum amplitude difference between the pulses. Purple arrows indicate polarization at the transmitter. b) Zoom-in of Fig.~\ref{fig:results1} including also the result with paddles stationary (system is free running). Black horizontal lines indicate estimated limits based on power variations and the signal amplitude polarization dependence.}
\label{fig:results2}
\end{figure}
  When the paddles are moved, polarization-dependent loss (PDL) increases the power fluctuations, and the estimated phase noise level due to power fluctuations rises slightly (dashed blue trace) but not to the level seen in the experiment with dual polarization (purple). 

What is believed to be limiting the PMD cancellation is the unbalanced nature of the used direct photodetection. PDL causes the relative amplitudes of the two interleaved pulse trains to vary (Fig.~\ref{fig:results2}a), which means that when microwaves are extracted from the pulse train, the two polarizations are not weighted equally. Without knowledge of the relation between polarization and loss (which can vary uncontrollably), the peak-to-peak phase drift of direct detection (DD), $\Delta\tau_{DD}$, is
\begin{equation}
      \Delta \tau_{DD} \leqslant \frac{\Delta \tau_{PMD}}{2} \frac{\Delta V}{\overline{V}},
\end{equation}
where $\Delta\tau_{PMD}$ is the peak-to-peak differential group delay due to PMD, $\Delta V$ is the photodiode signal variation and $\overline{V}$ the signal average. For instance, if all optical power would move from the fastest state of polarization to the slowest, then $\Delta V = 2\overline{V}$ and $\Delta \tau_{DD} = \Delta \tau_{PMD}$. In the present case, $\Delta V/\overline{V}\approx$~6.2\% and $\Delta\tau_{PMD}=300~$fs, which yields a worst case phase drift of $9$~fs due to PDL. When this is added to the "background" phase variation, which is about 5~fs peak-to-peak without polarization perturbations (Fig.~\ref{fig:results2}b), a worst-case performance of about 14~fs peak-to-peak could be expected. The increased power fluctuations during polarization perturbations would increase this by a few fs more. As seen in Fig.~\ref{fig:results2}b, the phase stays mostly within this boundary, which is marked by the black, horizontal lines.




The well-known AM-to-PM noise conversion problem in photodiodes has motivated the development of nonlinear \cite{Schibli2003a, Kim2007a} and linear \cite{Hou2015a, Wang2021a, Caldwell2022a, Wang2023a} detectors for balanced optical-optical phase detection, as also used here for the in-loop detector, and balanced optical-microwave phase detectors \cite{Kim2006a, Lessing2013a, Endo2018a, Jeon2018a, Ahn2022a}. In the present case with varying polarization, the detector {\color{black} in the receiver} should be double-balanced, that is, balanced in power as well as in polarization. In coherent communication it is standard practice to detect two orthogonal polarization components simultaneously and separately using dual-polarization optical hybrids. These have also been shown to be effective in preventing signal fading in long-haul optical frequency transfer \cite{Clivati2020a}.
 In principle, a similar solution should be straightforward to apply to the present case; however, such 
 a "double-balanced" detector has not, to the best of the authors' knowledge, been demonstrated yet.


\textbf{Conclusions.}
Polarization-switched pulse interleaving was shown to be simple and effective against PMD in a 30-km dispersion-compensated closed-loop setup relevant for precision time transfer. Phase drift due to differential group delay was reduced by more than an order of magnitude, from 300~fs to <~20~fs, limited by the polarization and power sensitivity of the out-of-loop phase detector that was based on direct photodetection. 
  To reach the sub-femtosecond regime over hundreds of kilometers of SM  fibre, drift-free "double-balanced" phase detectors that are insensitive to both power and polarization changes might be needed {\color{black}in the receiver.}  {\color{black} This work has highlighted PMD cancellation over time scales of 0.1-100s. It should be noted that only two consecutive pulses are needed, so the cancellation should in principle work up towards the fundamental repetition frequency of the laser.  Also and as shown, the technique works even when a single polarization is used for the in-loop phase stabilization, as is common practice; therefore, the long-term operation should not be any different from traditional systems except that polarization induced drift on any relevant time scale is also canceled as long as the internal pulse interleaver is made sufficiently stable.}




\begin{backmatter}
\bmsection{Funding} 
Research Council of Finland (decisions 339821, 328786 and 320168) and the European Metrology Programme for Innovation and Research (20FUN08 NEXTLASERS). 


\bmsection{Acknowledgments} The project 20FUN08 NEXTLASERS has received funding from the EMPIR programme cofinanced by the Participating States and from the European Union’s Horizon 2020 Research and Innovation Programme. The work is also part of the Research Council of Finland Flagship Programme, Photonics Research and Innovation (PREIN), decision 320168.


\smallskip
\bmsection{Disclosures} The authors declare no conflicts of interest.






\bigskip


\bmsection{Supplemental document}
None. 

\end{backmatter}

\bibliography{sample}

\begin{thebibliography}{10}
\newcommand{\enquote}[1]{``#1''}

\bibitem{Caldwell2023a}
E.~D. Caldwell, J.-D. Deschenes, J.~Ellis, \emph{et~al.}, {\protect\JournalTitle{Nature}} \textbf{618}, 721 (2023).

\bibitem{Xu2021a}
D.~Xu, O.~Lopez, A.~Amy-Klein, and P.-E. Pottie, {\protect\JournalTitle{Opt. Express}} \textbf{29}, 17476 (2021).

\bibitem{Zhang2017a}
H.~Zhang, G.~Wu, X.~Li, and J.~Chen, {\protect\JournalTitle{Metrologia}} \textbf{54}, 94 (2017).

\bibitem{Gibbon2015a}
T.~B. Gibbon, E.~K.~R. Kipnoo, R.~R.~G. Gamatham, \emph{et~al.}, {\protect\JournalTitle{Journal of Astronomical Telescopes, Instruments, and Systems}} \textbf{1}, 1  (2015).

\bibitem{Droste2013a}
S.~Droste, F.~Ozimek, T.~Udem, \emph{et~al.}, {\protect\JournalTitle{Phys. Rev. Lett.}} \textbf{111}, 110801 (2013).

\bibitem{Bercy2014a}
A.~Bercy, F.~Stefani, O.~Lopez, \emph{et~al.}, {\protect\JournalTitle{Phys. Rev. A}} \textbf{90}, 061802 (2014).

\bibitem{Guillou-Camargo2018a}
F.~Guillou-Camargo, V.~M\'{e}noret, E.~Cantin, \emph{et~al.}, {\protect\JournalTitle{Appl. Opt.}} \textbf{57}, 7203 (2018).

\bibitem{Schioppo2022a}
M.~Schioppo, J.~Kronj{\"a}ger, A.~Silva, \emph{et~al.}, {\protect\JournalTitle{Nature Communications}} \textbf{13}, 212 (2022).

\bibitem{Xie2011}
C.~Xie, {\protect\JournalTitle{Opt. Express}} \textbf{19}, B915 (2011).

\bibitem{Ning2005a}
T.~Ning, L.~Pei, Y.~Liu, \emph{et~al.}, {\protect\JournalTitle{Optica Applicata}} \textbf{35}, 277 (2005).

\bibitem{Xin2014a}
M.~Xin, K.~\c{S}afak, M.~Y. Peng, \emph{et~al.}, {\protect\JournalTitle{Opt. Express}} \textbf{22}, 14904 (2014).

\bibitem{Xin2017a}
M.~Xin, K.~Şafak, M.~Y. Peng, \emph{et~al.}, {\protect\JournalTitle{IEEE Journal of Selected Topics in Quantum Electronics}} \textbf{23}, 97 (2017).

\bibitem{Xin2018b}
M.~Xin, K.~\c{S}afak, and F.~X. K\"{a}rtner, {\protect\JournalTitle{Optica}} \textbf{5}, 1564 (2018).

\bibitem{He2013a}
Y.~He, B.~J. Orr, K.~G.~H. Baldwin, \emph{et~al.}, {\protect\JournalTitle{Opt. Express}} \textbf{21}, 18754 (2013).

\bibitem{Deschenes2016a}
J.-D. Desch\^enes, L.~C. Sinclair, F.~R. Giorgetta, \emph{et~al.}, {\protect\JournalTitle{Phys. Rev. X}} \textbf{6}, 021016 (2016).

\bibitem{Sinclair2019a}
L.~C. Sinclair, H.~Bergeron, W.~C. Swann, \emph{et~al.}, {\protect\JournalTitle{Phys. Rev. A}} \textbf{99}, 023844 (2019).

\bibitem{Abuduweili2020a}
A.~Abuduweili, X.~Chen, Z.~Chen, \emph{et~al.}, {\protect\JournalTitle{Opt. Express}} \textbf{28}, 39400 (2020).

\bibitem{Yao2002a}
X.~S. Yao, \enquote{Fiber devices based on fiber squeezer polarization controllers,} US Patent 6,493,474 (2002).

\bibitem{Kersey1990a}
A.~Kersey, M.~Marrone, and A.~Dandridge, {\protect\JournalTitle{Journal of Lightwave Technology}} \textbf{8}, 838 (1990).

\bibitem{Zhang2014a}
A.~Zhang, Y.~Dai, F.~Yin, \emph{et~al.}, {\protect\JournalTitle{Opt. Express}} \textbf{22}, 21560 (2014).

\bibitem{Fordell2022a}
T.~Fordell, {\protect\JournalTitle{Opt. Express}} \textbf{30}, 6311 (2022).

\bibitem{Kim2007a}
J.~Kim, J.~Chen, Z.~Zhang, \emph{et~al.}, {\protect\JournalTitle{Opt. Lett.}} \textbf{32}, 1044 (2007).

\bibitem{Kartner2007a}
F.~X. K\"{a}rtner, F.~N.~C. Wong, and J.-W. Kim, \enquote{Compact background-free balanced cross-correlators,}  (2007). US Patent 7,940,390 B2.

\bibitem{Callahan2014a}
P.~T. Callahan, K.~Safak, P.~Battle, \emph{et~al.}, {\protect\JournalTitle{Opt. Express}} \textbf{22}, 9749 (2014).

\bibitem{Safak2022a}
K.~\c{S}afak, A.~Dai, M.~Xin, \emph{et~al.}, \enquote{Extreme-timing-resolution with waveguide-based balanced optical cross-correlators,} in \emph{Conference on Lasers and Electro-Optics,}  (Optica Publishing Group, 2022), p. STh5N.3.

\bibitem{Taylor2011a}
J.~Taylor, S.~Datta, A.~Hati, \emph{et~al.}, {\protect\JournalTitle{IEEE Photonics Journal}} \textbf{3}, 140 (2011).

\bibitem{Zhang2012a}
W.~Zhang, T.~Li, M.~Lours, \emph{et~al.}, {\protect\JournalTitle{Applied Physics B}} \textbf{106}, 301 (2012).

\bibitem{Sun2014a}
W.~Sun, F.~Quinlan, T.~M. Fortier, \emph{et~al.}, {\protect\JournalTitle{Phys. Rev. Lett.}} \textbf{113}, 203901 (2014).

\bibitem{Schibli2003a}
T.~R. Schibli, J.~Kim, O.~Kuzucu, \emph{et~al.}, {\protect\JournalTitle{Opt. Lett.}} \textbf{28}, 947 (2003).

\bibitem{Hou2015a}
D.~Hou, C.-C. Lee, Z.~Yang, and T.~R. Schibli, {\protect\JournalTitle{Opt. Lett.}} \textbf{40}, 2985 (2015).

\bibitem{Wang2021a}
T.~Wang, Q.~Ren, K.~\c{S}afak, \emph{et~al.}, {\protect\JournalTitle{Opt. Express}} \textbf{29}, 38140 (2021).

\bibitem{Caldwell2022a}
E.~D. Caldwell, L.~C. Sinclair, N.~R. Newbury, and J.-D. Deschenes, {\protect\JournalTitle{Nature}} \textbf{610}, 667 (2022).

\bibitem{Wang2023a}
T.~Wang, M.~Li, Y.~Zhang, and M.~Xin, {\protect\JournalTitle{Opt. Lett.}} \textbf{48}, 5201 (2023).

\bibitem{Kim2006a}
J.~Kim, F.~X. K\"{a}rtner, and F.~Ludwig, {\protect\JournalTitle{Opt. Lett.}} \textbf{31}, 3659 (2006).

\bibitem{Lessing2013a}
M.~Lessing, H.~S. Margolis, C.~T.~A. Brown, \emph{et~al.}, {\protect\JournalTitle{Opt. Express}} \textbf{21}, 27057 (2013).

\bibitem{Endo2018a}
M.~Endo, T.~D. Shoji, and T.~R. Schibli, {\protect\JournalTitle{Scientific Reports}} \textbf{8}, 4388 (2018).

\bibitem{Jeon2018a}
C.-G. Jeon, Y.~Na, B.-W. Lee, and J.~Kim, {\protect\JournalTitle{Opt. Lett.}} \textbf{43}, 3997 (2018).

\bibitem{Ahn2022a}
C.~Ahn, Y.~Na, M.~Hyun, \emph{et~al.}, {\protect\JournalTitle{Photon. Res.}} \textbf{10}, 365 (2022).

\bibitem{Clivati2020a}
C.~Clivati, P.~Savio, S.~Abrate, \emph{et~al.}, {\protect\JournalTitle{Opt. Express}} \textbf{28}, 8494 (2020).

\end{thebibliography}



\ifthenelse{\equal{\journalref}{aop}}{%
\section*{Author Biographies}
\begingroup
\setlength\intextsep{0pt}
\begin{minipage}[t][6.3cm][t]{1.0\textwidth} 
  \begin{wrapfigure}{L}{0.25\textwidth}
    \includegraphics[width=0.25\textwidth]{john_smith.eps}
  \end{wrapfigure}
  \noindent
  {\bfseries John Smith} received his BSc (Mathematics) in 2000 from The University of Maryland. His research interests include lasers and optics.
\end{minipage}
\begin{minipage}{1.0\textwidth}
  \begin{wrapfigure}{L}{0.25\textwidth}
    \includegraphics[width=0.25\textwidth]{alice_smith.eps}
  \end{wrapfigure}
  \noindent
  {\bfseries Alice Smith} also received her BSc (Mathematics) in 2000 from The University of Maryland. Her research interests also include lasers and optics.
\end{minipage}
\endgroup
}{}

\end{document}